# Supergranulation Velocity Field from the MDI (SOHO) Data

L. D. Parfinenko, V. I. Efremov, and A. A. Solov'ev

*Main (Pulkovo) Astronomical Observatory, Russian Academy of Sciences, St. Petersburg, Russia*

*e-mail: parfinenko@mail.ru*

**Abstract**—Long (up to 100 hours) time series of *SOHO/MDI* Doppler data are analyzed. The power spectrum of radial velocity time series in the quiet photosphere is observed to have, along with the known 5-min mode, a stable strong mode with a period of about 32 h, which is close to the supergranulation cell lifetime. The spatial distribution of the amplitudes of these oscillations also coincides with the characteristic size scale of supergranulation (~35 Mm).

**DOI:** 10.1134/S0016793214080143

## 1. INTRODUCTION

In the study of long-period oscillations in sunspot magnetic field and radial velocities, we showed the following:

(1) Long-period sunspot oscillations have a multimode character: oscillations are confidently detected in the bands with periods of 60–80, 135–170, and 220–240 min, with the oscillation power decreasing monotonically with decreasing period.

(2) The limiting (lowest frequency) mode of the sunspot magnetic field natural oscillations is the $M1$ mode, with periods ranging from 12 to 28–32 h. This mode is the most regular; it exists stably in sunspots throughout the observation period (5–7 days). However, the spatial distribution of its amplitude shows quasi-periodic ascents separated by time intervals of about 1.5–2 days, which is close to the average lifetime of supergranulation cells. The oscillation amplitude at this low-frequency mode is 200–250 G for magnetic field and 60–110 m/s for radial velocity. The limiting mode period is significantly and nonlinearly dependent on the magnitude of the sunspot magnetic field. As follows from our previous studies (Efremov et al., 2010; 2012; 2014) and the shallow sunspot model (Solov'ev and Kirichek, 2008; 2009), the period of the natural mode $M1$ decreases with increasing field strength, reaching a minimum value of ~10–12 h at around 2600–2700 G and then rises again in the range 2700 to 3400 G. At field strengths of 2600–2700 G, the sunspot is the most stable and has a minimum natural period (Efremov et al., 2014, Fig. 6).

(3) The wavelet character of the amplitude oscillations for the limiting mode $M1$ leads to the sunspot oscillation power spectrum having an even lower frequency mode $M2$, with a period of about 30–48 h. $M2$ is not a natural mode because, as it turned out, its period does not depend on the sunspot magnetic field and its amplitude may sometimes be less than that of $M1$. The $M2$ mode reflects, we believe, the quasi-period of an external perturbing force attributed to the sunspot dynamic perturbations from the surrounding supergranulation cells.

The aim of this work is to detect in the quiet photosphere oscillations with a period close to that of the $M2$ mode. To this end, we investigated the low-frequency power spectrum of radial velocity time series for the quiet photosphere.

## 2. OBSERVATION DATA AND PROCESSING TECHNIQUE

The observation data used in the study were long (up to 100 h) series of FITS magnetograms, Dopplergrams, and intensitygrams obtained using the *Michelson Doppler Image* (*MDI*) on board the *Solar and Heliospheric Observatory* (SOHO) (Scherrer et al., 1995). We used observation series recorded at 1-min cadence. The *SOHO/MDI* space data includes a time series of full solar disk Dopplergrams (http://soi.stanford.edu/data/). In each Dopplergram we selected a radial velocity value strictly for one and the same point in the photosphere throughout the whole multiday observation series. In order to construct these time series, it was necessary to stabilize the image of an object (spot, pore, etc.) moving along the Sun's surface. To this end, we used magnetograms or intensitygrams of stable spots having a regular round shape without bridges. The images were stabilized using an extreme reference for a distributed object that had been immersed in a *strip*, a limited fixed frame. After the first run, we obtained a pair of coordinate functions $\{X(n), Y(n)\}$ to create a scenario describing the motion of the reference (in our case, extreme) in the strip during the observation period. Immersing the spot in a small frame and making the frame boundaries move according to the scenario, we stabilized the object. Then we could construct the time series for a given parameter at any point in or near the spot. The processing technique is described in (Efremov et al. 2010; 2012; 2014). Due to the low degree of structure and low contrast of the Dopplergrams, pegging the results to the extreme reference was very difficult even within a spot, to say nothing of the quiet photosphere.

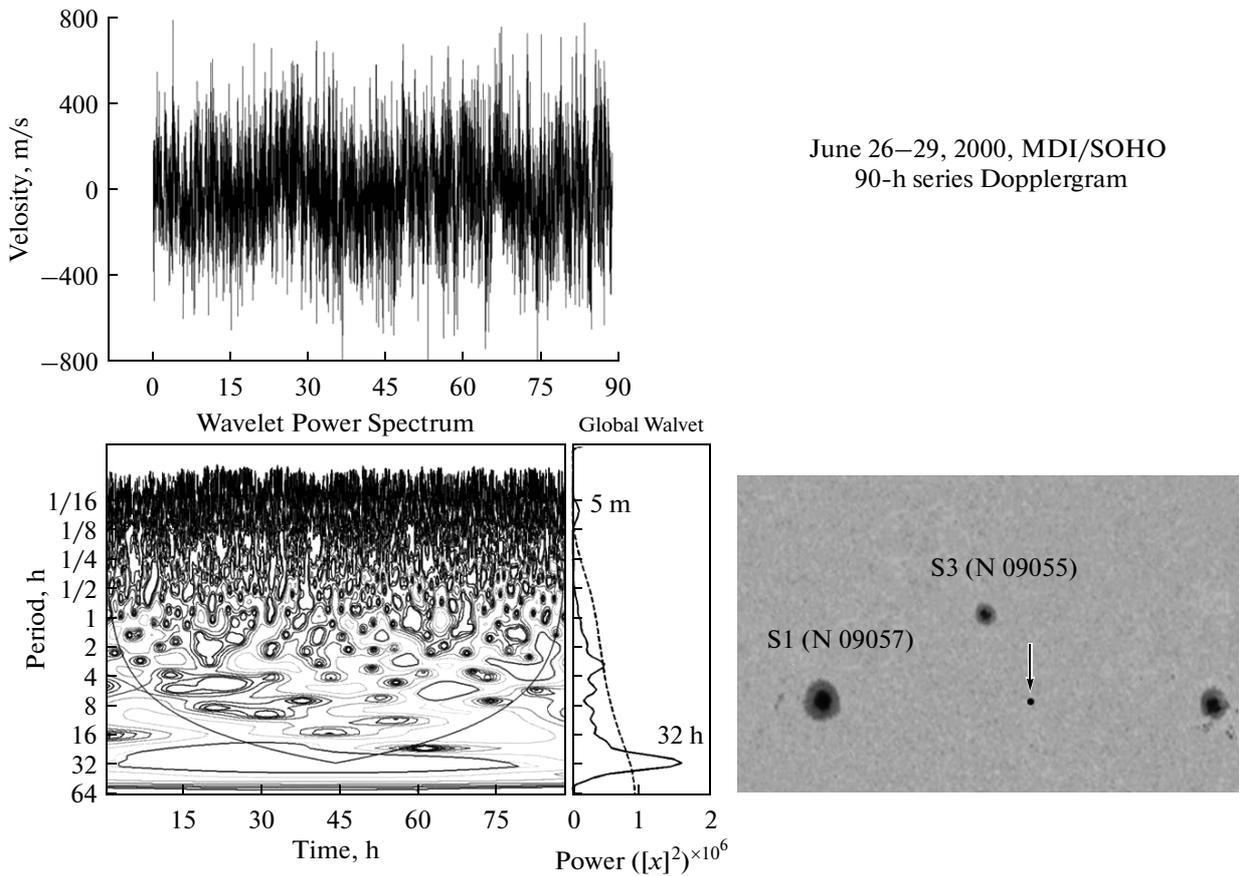

**Fig. 1.** The upper left panel shows the radial velocity time series at the photosphere point marked by an arrow (right panel). The lower left panel shows the time series wavelet.

Hence, we used, along with the Dopplergrams, a synchronous series of magnetograms or intensitygrams, which allow a relatively simple and reliable sunspot stabilization. The simultaneous observation should provide the magnitudes of the radial velocity $V_r$ and magnetic field $H$ (or continuum intensity $I_c$), which can be quite confidently pegged to the extreme reference near the sunspot center. Then, synchronizing the data and using the sunspot motion scenario for the magnetic field ($H$) or intensity ($I_c$), we applied the scenario to the Dopplergrams ($V_r$). Obviously, the resulting scenario can be used not only for the spot but also for any point in the quiet photosphere moving across the disk at the latitude of the selected spot in the same time interval.

## 3. RESULTS AND DISCUSSION

Our assumption was that the $M2$ mode in the sunspot magnetic field oscillations is attributed to quasi-periodic impacts of an external perturbing force with a sufficiently large amplitude due to the "shocks" from supergranulation cells, which arise and decay with a characteristic period of about 30–40 h. Supergranulation, which was discovered by Hart (1954) and studied in detail by Leighton et al. (1962), is known to have the following parameters: a horizontal size of ~30–35 Mm, horizontal velocity of 200–500 m/s; and velocity of 50–100 m/s for the upflows at the cell center and 100–200 m/s for the downflows at the boundaries. The lifetime of most cells is 15–30 h, sometimes two days or more. It should be noted that there is some evidence of wave-like properties of supergranulation (Wolff, 1995; Gizon et al., 2003; Schou, 2003). We studied the wave-like and oscillation properties of supergranulation from radial velocity variations at a fixed point in the photosphere by measuring directly the radial velocities of the emitting gas. A complicating factor is that Doppler shifts in supergranulation have a weak vertical component, i.e., the plasma spreads out from the cell center mainly in the horizontal direction.

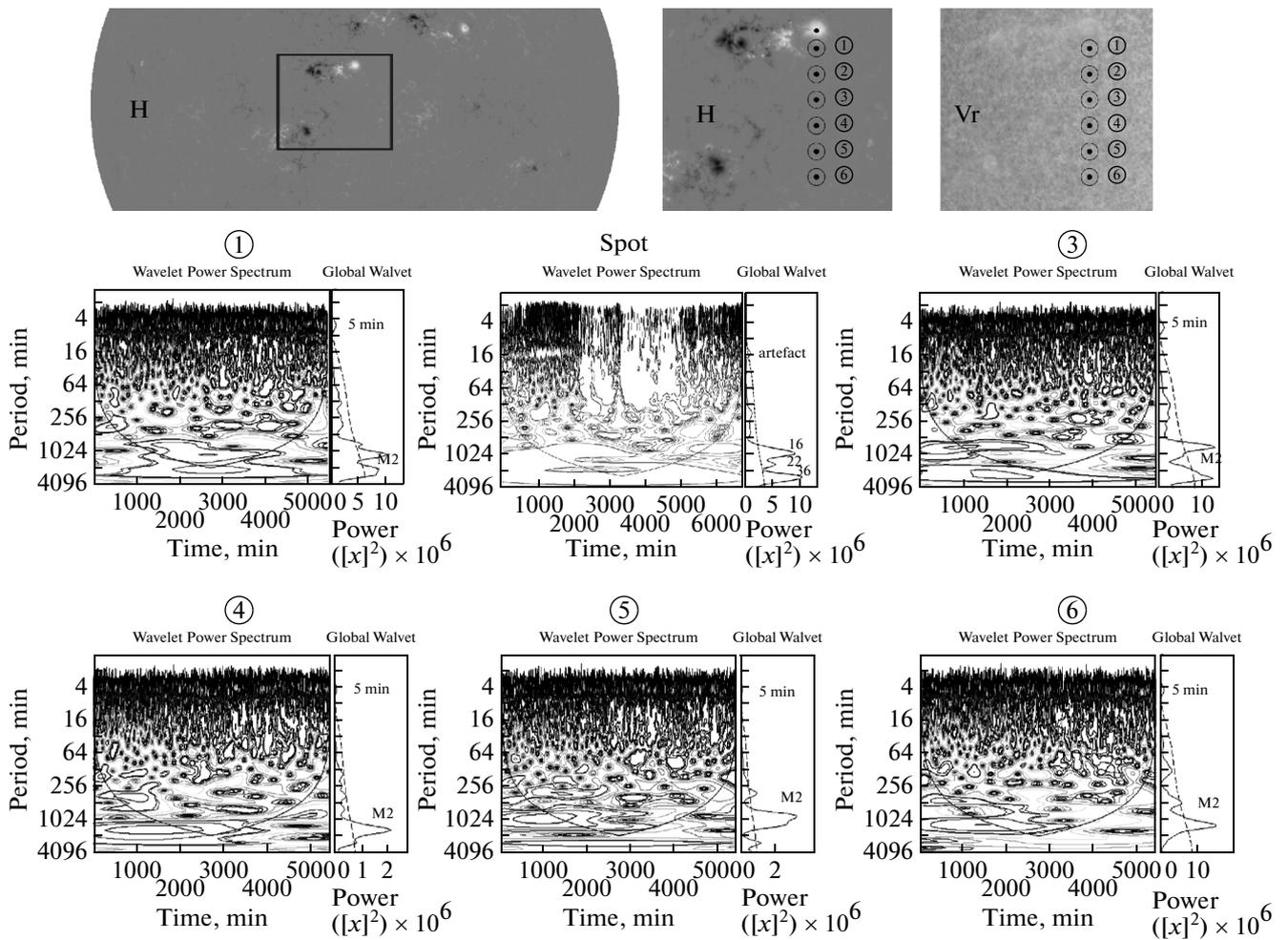

**Fig. 2.** The upper panel shows the position of the analyzed fragment in the full solar disk magnetogram. In the middle and on the right is shown, on a larger scale, the analyzed fragment of the magnetogram and Dopplergram. The points are indicated at which velocity oscillations were analyzed: six points in the photosphere and one point at the spot center. The middle and lower panels present the corresponding wavelets for the radial velocity time series for points *1*, *3*, *4*, *5*, and *6* in the photosphere and the central point of the spot.

To investigate the oscillations, we applied wavelet analysis (Torrence et al., 1998), using a fifth order Morlet wavelet as an analyzing function. Trends in the radial velocity time series were approximated with fourth-order polynomials.

Preliminary studies were conducted for observations made in the period from June 26 to June 29, 2000 (SOHO/MDI 90-hour series). Figure 1 shows the radial velocity time series at an arbitrary point in the quiet photosphere (marked by an arrow) and the corresponding wavelet. The power spectrum of the radial velocity time series in the quiet photosphere turned out to have, along with the known 5-min mode (Leibacher and Stein, 1971), a strong low-frequency mode with a period of about 33 h.

For a more detailed study, we selected the 100-h SOHO/MDI series for the period from March 31 to April 4, 2002 in the area of NOAA 09887.

The power spectra of radial velocity oscillations at arbitrarily selected points in the quiet photosphere (Fig. 2) are observed to have a strong low-frequency mode with a period of about 30 h, which is close to that of $M2$. The wavelet for the spot center shows no 5-min mode but is observed to have a technical artifact (a ~16-min mode). The latter was described by Efremov et al. (2010; 2012). The spot image moves across the matrix due to the Sun's rotation. The artifact appears when the maximum field strength point, or maximum velocity point, moves from one pixel to another in the matrix because of a large magnetic field or velocity gradient in the spot (produces a short jump in the signal). In the quiet photosphere, the velocity gradient is small and the artifact does not manifest itself.

Our next step was to study the spatial distribution of the $M2$ mode in the directions along the meridian and along the parallel. To this end, we searched the Dopplergrams for the corresponding scans for a quiet area in the photosphere in the vicinity of the sunspot. Then, collecting a sequence of scans for the whole observation, we first built an $L-T$ map, which is the timebase of the radial velocities in each selected scan, and then used the resulting $L-T$ map to build $\Omega-L$ maps. In these maps the $Y$ axis plots the oscillation frequency and the $X$ axis plots the scan number. This is a highly visual representation of the process, ensuring its readability: we see both the spatial distribution of the process power, i.e., the localization of the power in space, and the temporal distribution of the process, i.e., its characteristic periodicity (Efremov et al., 2007).

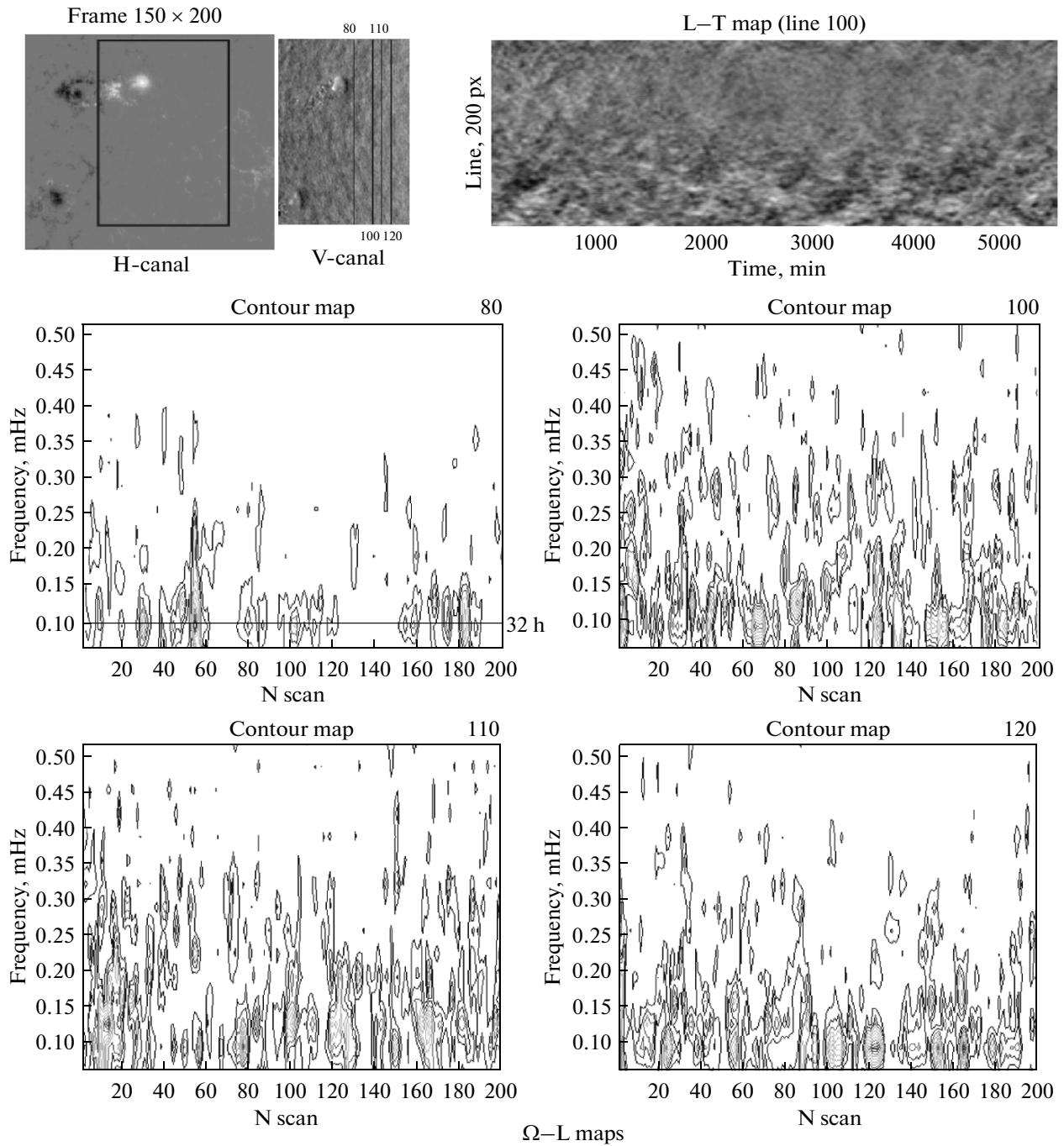

**Fig. 3.** The upper panel shows a fragment of the magnetogram and Dopplergram with the meridional sections used to analyze the spatial distribution of velocity oscillations. On the right is shown the $L-T$ map for the 100th section. The middle and lower panels present the $\Omega-L$ maps for 80, 100, 110, and 120 sections.

Figures 3 and 4 show the results of our study for the direction along the meridian and parallel. It is clear that the oscillation power can be grouped into clusters

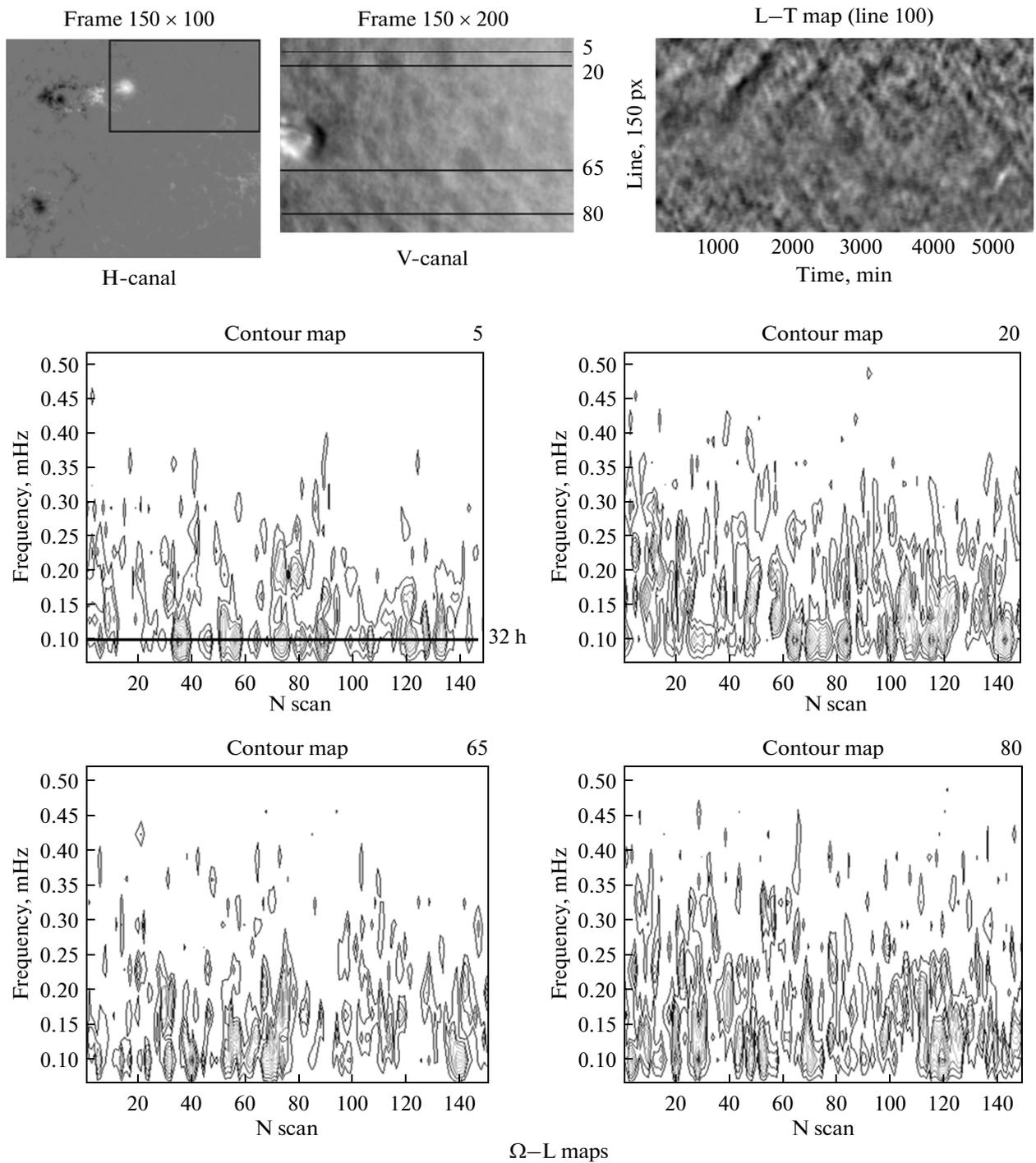

**Fig. 4.** The upper panel shows a fragment of the magnetogram and Dopplergram with the sections made along the parallel to analyze the spatial distribution of velocity oscillations. On the right is shown the $L-T$ map for the 20th section. The middle and lower panels present the $\Omega-L$ maps for 2, 20, 65, and 80 sections.

at a period of about 32 h, with the distance between the neighboring clusters being about 35 Mm.

These results allowed us to the estimate the height $h$ of the supergranulation cell. Taking into account the continuity of flows in the convective cell, the average time of gas lift in the cell may be assumed to be equal to that of its horizontal flow: $R/\tilde{V}_{rad} \approx h/\tilde{V}_{lift}$, where $R$ is the radius of the convective cell; $h$ is its characteristic thickness (height); and $\tilde{V}_{rad}$ and $\tilde{V}_{lift}$ are the average velocities of the radial flow and gas lift, respectively. Based on the data obtained, we can assume that the characteristic gas lift velocity is $\tilde{V}_{lift} = 100$ m/s and take the known values of 200 m/s to 400 m/s for the average horizontal gas flow velocity. Then we have $R \approx (2-5)h$. Since the radius $R$ of the supergranulation cell cross-section is on average ~17.5 Mm, then the height of the cell is $h = (3-8)$ Mm; i.e., supergranulation cells prove to be planar formations occurring in the upper layers of the convective zone. This is independent evidence supporting the hypothesis that supergranulation is induced by convective instability in the area of He+ recombination at a depth of (2–8) Mm (Simon and Leighton, 1964). However, it might be possible that the observed supergranulation pattern is due to the superimposition of wave and oscillation processes upon the hydrodynamics of convective transport.

It should be noted that in this study we did not use data from *SDO/HMI*, which has a better spatial resolution, because the data for the target frequency range may have a daily Doppler artifact attributed to this instrument. (The *SDO/HMI* data can be used to study long-period sunspot oscillations if the sunspot magnetic field intensity is less than 2000 G, at which there are no apparent manifestations of the artifact (Smirnova et al., 2013)). As to the *SOHO/MDI* instrument, we showed previously (Efremov et al., 2010; 2014) that the observations series recorded at 1-min cadence in the target low-frequency spectral region have no significant artifacts.

## 4. CONCLUSIONS

(1) The quiet photosphere velocity field is observed to have vertical plasma oscillations with a characteristic period of $T \sim 32$ h.

(2) The oscillation power is distributed across the photosphere as clusters with a characteristic spatial scale of $L \sim 35$ Mm, both along the meridian and along the parallel.

(3) The amplitude of radial velocity oscillations in these clusters is approximately 100 m/s.

(4) The oscillation clusters discovered in the magnetic field-free photosphere have the same spatial and temporal characteristics as supergranulation cells, which is substantial evidence for the physical identification of these phenomena.

(5) Presumably, it is the motions of superconvective cells that induce $M2$ mode with a characteristic period of $T \sim 32$ h in the power spectra of the sunspot magnetic field and radial velocity oscillations.

## ACKNOWLEDGMENTS

We thank Dr. P.H. Scherrer and the *SOHO* team for the opportunity to use the *SOHO/MDI* observation data.

This work was supported by the Presidium of the Russian Academy of Sciences, project nos. P-21 and P-22; the Science School Support Program, project no. NSh-1625.2012.2, and the Russian Foundation for Basic Research, project no. 13-02-00714.

*Translated by A. Kobkova*